\documentclass[prb,twocolumn,showpacs,preprintnumbers,amsmath,amssymb,floats,aps,10pt]{revtex4}
\usepackage{graphicx}
\usepackage{amsmath}
\usepackage{epstopdf}
\usepackage{amsbsy}
\usepackage{color}
\usepackage[font=small,justification=raggedright]{caption,subfig}
\usepackage{dcolumn}
\usepackage{bm}

\newcommand{\bmx}{\begin{bmatrix}}
\newcommand{\emx}{\end{bmatrix}}
\newcommand{\bpmx}{\begin{pmatrix}}
\newcommand{\epmx}{\end{pmatrix}}
\newcommand{\beq}{\begin{equation}}
\newcommand{\eeq}{\end{equation}}

\renewcommand{\H}{\mathcal{H}}
\renewcommand{\k}{\boldsymbol{\mathrm{k}}}

\newcommand{\q}{\boldsymbol{\mathrm{q}}}

\renewcommand{\r}{\boldsymbol{\mathrm{r}}}
\newcommand{\up}{\uparrow}
\newcommand{\down}{\downarrow}

\begin{document}

\title{Signatures of orbital loop currents in the spatially resolved local density of states}
\author{W. H. P. Nielsen$^1$, W. A. Atkinson$^2$, and B. M. Andersen$^1$}
\affiliation{$^1$Niels Bohr Institute, University of Copenhagen, Universitetsparken 5, DK-2100 Copenhagen,
Denmark\\
$^2$Department of Physics and Astronomy, Trent University, Peterborough, Ontario, Canada K9J 7B8}

\date{\today}

\begin{abstract}

Polarized neutron scattering measurements have suggested that intra-unit cell antiferromagnetism may be associated with the pseudogap phase. Assuming that loop current order is responsible for the observed magnetism, we calculate some signatures of such circulating currents in the local density of states around a single non-magnetic impurity in a coexistence phase with superconductivity. We find a distinct C$_4$ symmetry breaking near the disorder which is also detectable in the resulting quasi-particle interference patterns. 

\end{abstract}

\pacs{74.20.-z, 74.25.Jb, 74.50.+r, 74.72.-h}

\maketitle

\section{Introduction}

The understanding of the pseudogap phase in cuprate materials constitutes an outstanding challenge in condensed matter physics. At present, it remains unsolved whether the phase is caused by singlet formation naturally exhibited in strong-coupling models, or rather by an elusive spontaneously symmetry-broken phase.  

In a recent series of spin-polarized neutron scattering experiments, it has been observed that intra-unit cell antiferromagnetic order sets in at $T^*$, the pseudogap temperature in the underdoped regime of YBa$_2$Cu$_3$O$_{6+x}$ and HgBa$_2$CuO$_{4+\delta}$.\cite{fauque,mook,li,li2} Recently, a short-range magnetic intra-unit cell signal was also identified in La$_{1.915}$Sr$_{0.085}$CuO$_{4}$ though at temperatures significantly below $T^*$.\cite{baledent} These results suggest that the pseudogap phase may in fact be associated with a true phase transition with an associated spontaneously broken symmetry. Varma has predicted that the pseudogap phase is caused by equilibrium loop currents which break the time-reversal symmetry but preserve the translational invariance of the CuO$_2$ lattice.\cite{varma,varma2} A recent theoretical study\cite{weber} found that such loop currents can be stabilized by using the apical oxygen ions leading to ordered moments rotated away from the axis perpendicular to the CuO$_2$ plane in qualitative agreement with the neutron measurements.  

The possible existence of loop order remains highly controversial, however, since local probes fail to detect the weak magnetic field associated with such orbital currents. For example, nuclear magnetic resonance (NMR), nuclear quadropole resonance (NQR), and muon spin relaxation ($\mu$SR) have failed to detect any onset of magnetic order setting in at $T^*$.\cite{straassle_old,macdougall,sonier,straassle} Most recently, a high-precision zero-field $\mu$SR study on La$_{2-x}$Sr$_x$CuO$_4$ found no evidence of static magnetic order in the pseudogap phase at doping levels above $x=0.13$, which is outside the spin-glass phase in LSCO.\cite{huang} Various proposals, including minority phases, loop-order fluctuations, and local muon destruction of loop order have surfaced to reconcile the 
apparent contradiction between local probes and neutron experiments.\cite{shekter,millis}
At present, the possible existence of loop order remains unsettled and therefore new proposals for its detection seem desirable. 

Scanning tunneling spectroscopy (STS) is a natural approach to take.  STS experiments
on $\mathrm{Bi_2Sr_2CaCu_2O_8}$, for example, have found evidence for intra-unit cell ``nematic'' symmetry breaking, which shows up as an inequivalency of oxygen $\mathrm{p}_x$ and $\mathrm{p}_y$ orbitals.\cite{lawler}  Unfortunately, loop order does not have such a simple experimental signature, since the local density of states (LDOS) has the same $C_4$ symmetry as the lattice. The loop currents are expected to generate a nonzero density of states at the Fermi energy,\cite{berg} but are not unique in this; other mechanisms, such as disorder, also generate low energy excitations.

In this paper, we show that there is a clear signature of loop currents if one considers
the LDOS near an isolated impurity.
In the literature, disorder effects have been previously suggested as a probe of the so-called $d$-density wave candidate for the pseudogap phase.\cite{chakravarty,wang,morr,andersenddw} Here, our aim is not to answer whether loop currents exist, but rather to suggest new ``smoking gun" experiments based on their assumed existence. We focus on the low-temperature coexistence phase of loop order and $d$-wave superconductivity since this is where 
STS measurements are most readily performed. 
We use a mean-field three-band model to study the effects of loop order on the LDOS near non-magnetic impurities. 
Our main result [Fig.~\ref{fig:LDOSpatterns}] is that there is a distinct $C_4$ symmetry breaking in the vicinity of the impurity, 
and an associated splitting of the so-called ``octet" vectors in the Fourier transformed scanning tunneling spectroscopy (FT-STS) patterns arising from quasi-particle interference (QPI). These effects can be explained solely from symmetry arguments, and result from momentum-selective shifts of the nodal Dirac points in the $d$-wave superconductor in the presence of loop order. 

\section{Model}
We employ the three-band Hubbard Hamiltonian with mean-field decoupled interactions.\cite{varma2} As shown recently, loop currents are stable in this model when the nearest-neighbor repulsion $V_{pd}$ is sufficiently large compared to the onsite repulsions on the copper $U_d$ and oxygen $U_p$ sites.\cite{fischer} In this paper, we assume the existence of loop currents and study their detectable consequences.
The three-band Hamiltonian is a minimal model containing the essential symmetries of the problem.
Experiments\cite{fauque,mook,li,li2} suggest that the current loops involve out-of-plane orbitals,\cite{weber} which are not part of the model. These additional orbitals are neglected for simplicity, as they do not change the rotational symmetries of the Hamiltonian,
and it is these symmetries that determine the symmetry of the LDOS pattern near an impurity.

In the three-orbital basis $\psi^\dagger(\k)=(d^\dagger_{\k},p^\dagger_{x,\k},p^\dagger_{y,\k})$ the normal-state Hamiltonian including orbital currents takes the form 
\begin{equation} \label{eq:Hamiltonian}
 \H=\sum_{\k} \psi^\dagger(\k) \big(\underbrace{H^0(\k)+H^R(\k)}_{H(\k)} \big) \psi(\k), 
\end{equation}
where $H^0(\k)$ and $H^R(\k)$ are $3\times 3$-matrices. $H^0(\k)$ describes the hopping part of the Hamiltonian as well as the decoupled on-site interactions, whereas $H^R(\k)$ carries the decoupled oxygen-copper interactions and is thus responsible for the loop currents. 

Writing out equation \eqref{eq:Hamiltonian} yields the following expression for the Hamiltonian:
\begin{widetext}
\begin{equation} \label{eq:3by3Hamiltonian}
 \H=\sum_{\k} \psi^\dagger(\k) \underbrace{\bmx \epsilon_d & 2it_{pd}s_x-Rc_x & -2it_{pd}s_y-Rc_y \\ -2it_{pd}s_x-R^*c_x & \epsilon_p & 4t_{pp}s_x s_y \\ 2it_{pd}s_y-R^*c_y & 4t_{pp}s_x s_y & \epsilon_p \emx}_{H(\k)} \psi(\k),
\end{equation}
\end{widetext}
where $c_i=\cos(k_i/2)$ and $s_i=\sin(k_i/2)$. Following
e.g.\ Hybertsen \emph{et al.},\cite{hybertsen} we use the following set of parameter values; $t_{pd}=-1.3$,
$t_{pp}=0.5t_{pd}$, $\epsilon_d=-1.5$, $\epsilon_p=-5.0$ (all in
eV). In Eq.(\ref{eq:3by3Hamiltonian}), $R$ denotes the mean-field
order parameter for the orbital currents, and is set to either zero
or, in the current-carrying case, to $0.1i$. This phase choice of $R$
leads to a current-carrying state in accordance with the original
mean-field formulation by Varma.\cite{varma2} The main results of this paper are not sensitive to the amplitude of $R$.\cite{footnote}

To include $d$-wave superconductivity, we change to a six-operator Nambu basis given by $\psi(\k)=( d^\dagger_{\k\up},p^\dagger_{x,\k\up},p^\dagger_{y,\k\up},d_{-\k\down},p_{x,-\k\down},p_{y,-\k\down})$. The Hamiltonian is then given by
\begin{equation} \label{eq:FullHamiltonian}
 \H=\sum_{\k}\psi^\dagger(\k) \bmx H(\k) & \Delta(\k) \\ \Delta^\dagger(\k) & -H^*(-\k) \emx \psi(\k),
\end{equation}
where $\Delta(\k)$ is the $3\times 3$ matrix describing the Cooper-pairing in our model. For simplicity, we take all Cooper pairing to be on the copper orbital which in the present case is a good approximation because the band crossing the Fermi level is mainly of Cu orbital character when $\epsilon_d=-1.5$, $\epsilon_p=-5.0$. We assume the gap function to have $d_{x^2-y^2}$-symmetry with
\begin{equation}
 \Delta(\k)=\Delta_0(\cos(k_x)-\cos(k_y))\cdot\text{diag}(1,0,0).
\end{equation}
Other choices for $\Delta(\k)$, involving pairing in the oxygen orbitals, are possible; however, these will not change the symmetry of the LDOS pattern.
To allow us to focus on spectral features below the gap, we take an artificially large $\Delta_0=0.25$ eV. 

The free Green's function for this system is now, for each $\k$, a $6\times 6$ matrix in orbital and spin indices given by
\begin{equation}
 G^0(\k,\omega)=(1(\omega+i\eta)-H'(\k))^{-1},
\end{equation}
where $H'(\k)$ is the $6\times 6$ matrix of Eq.\eqref{eq:FullHamiltonian}, and $\eta$ is an infinitesimal regulator. 

Finally, we introduce a pointlike non-magnetic impurity at the copper site belonging to the unit cell at
$\r_0$.
The impurity Hamiltonian is
\begin{equation}
 \H_{\text{imp}}=V_{\text{imp}}\cdot(\text{diag}(1,0,0)\oplus\text{diag}(-1,0,0))\delta(\r-\r_0).
\end{equation}
Using a standard $T$-matrix formalism, the full Green's function is then given by
\begin{equation}\label{Gfull}
 G(\r,\r',\omega)=G^0(\r-\r',\omega)+G^0(\r-\r_0,\omega)T(\omega)G^0(\r_0-\r',\omega),
\end{equation}
where
\begin{equation}
 T(\omega)=(1-H_{\text{imp}}G^0(0,\omega))^{-1}H_{\text{imp}}.
\end{equation}
From the full Green's function, the local density of states $\rho_{\ell}(\r,\omega)$ on the orbital $\ell$ in the unit cell $\r$ is readily obtained from the formula  (for the superconducting state)
\begin{equation}
  \rho_{\ell}(\r,\omega)=-\mbox{Im} \sum_{\sigma=\pm} G_{\ell\sigma,\ell\sigma}(\r,\r,\sigma\omega)/\pi,
\end{equation} 
with $\sigma$ labeling spin. The QPI spectrum is then obtained
from the Fourier transformed density of states:
\begin{equation}
\rho(\q,\omega) = \sum_{\r} e^{i\q\cdot\r} \left[ \rho_1(\r,\omega)
+ \rho_2(\r,\omega) e^{iq_x/2} + \rho_3(\r,\omega) e^{iq_y/2} \right ].
\end{equation}

\section{Results}

\begin{figure}[b]
\hspace{-0.05\columnwidth}
\subfloat[]{\includegraphics[width=0.2\columnwidth]{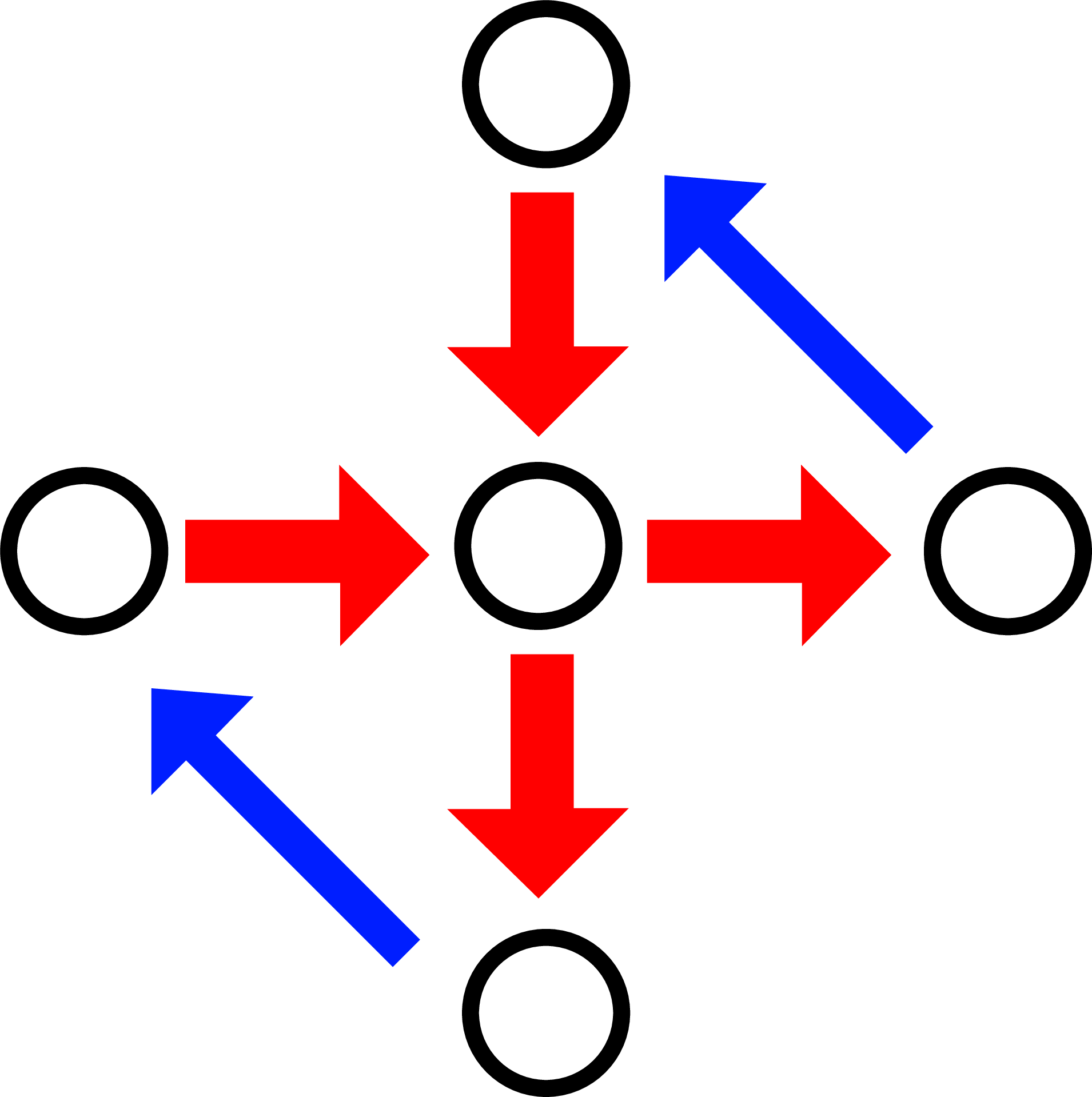}}%
\hspace{0.14\columnwidth}%
\subfloat[]{\includegraphics[width=0.2\columnwidth]{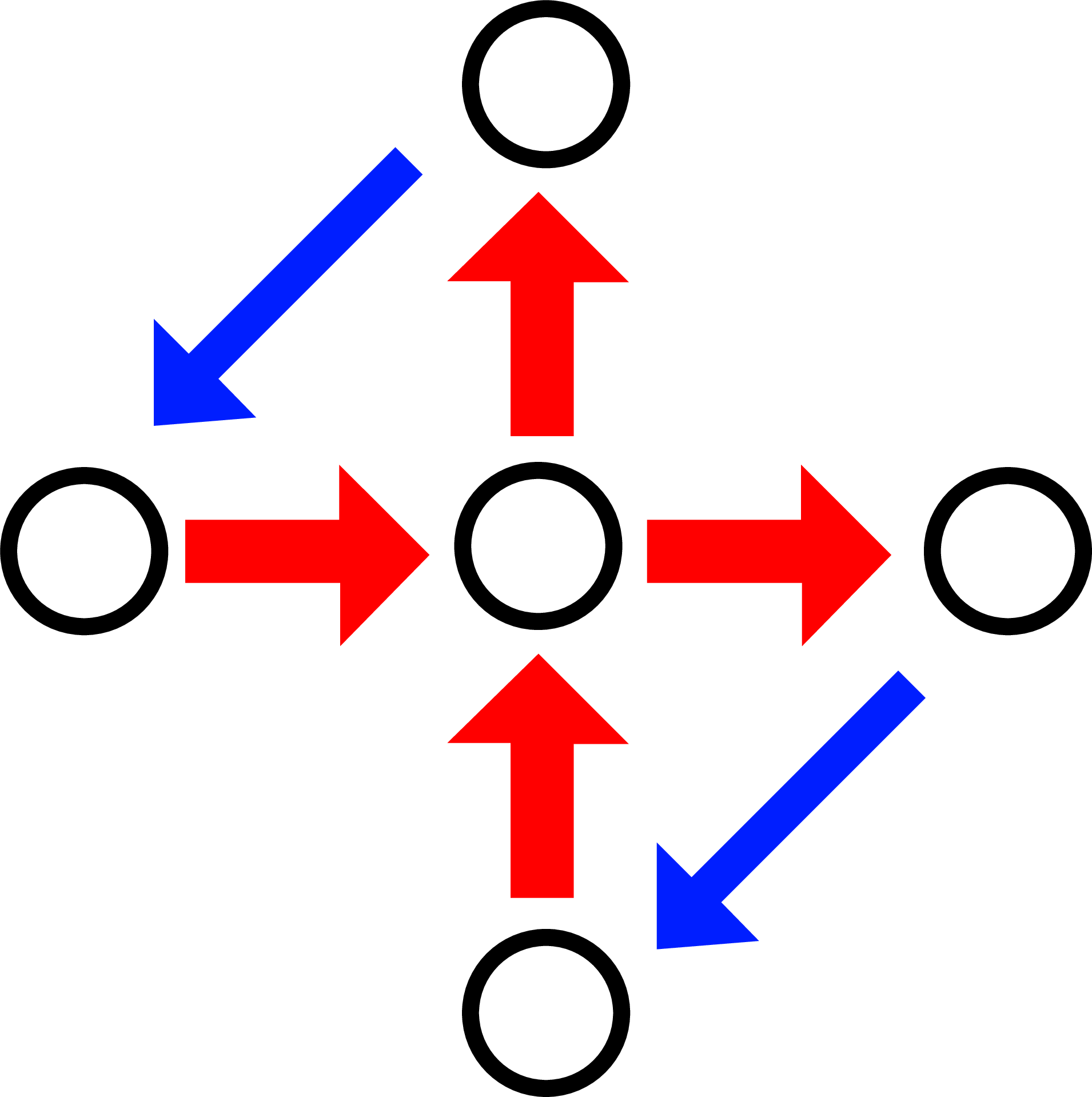}}%
\hspace{0.14\columnwidth}%
\subfloat[]{\includegraphics[width=0.2\columnwidth]{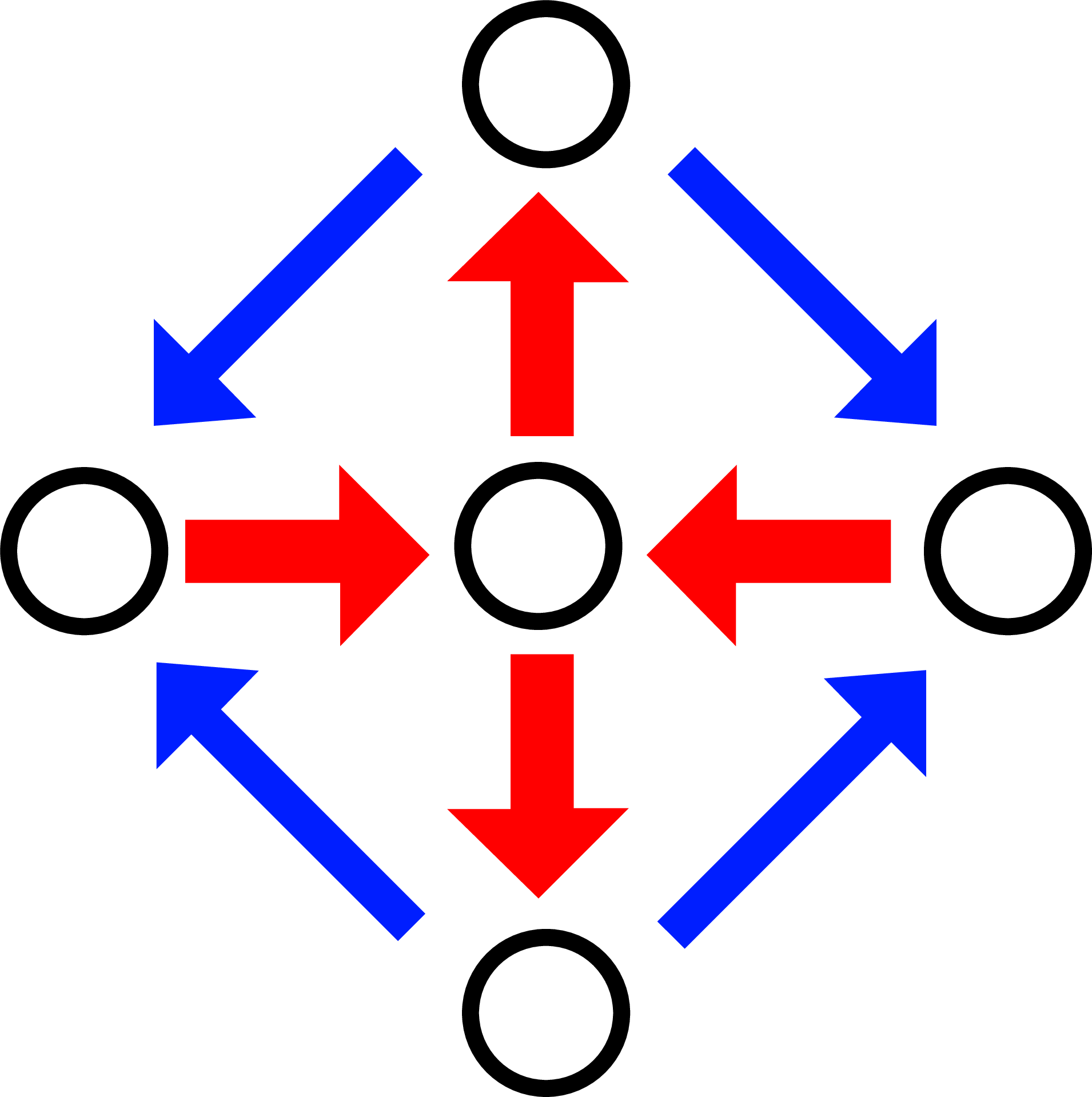}}\\%
\subfloat[\textcolor{white}{h}]{\includegraphics[width=0.3\columnwidth]{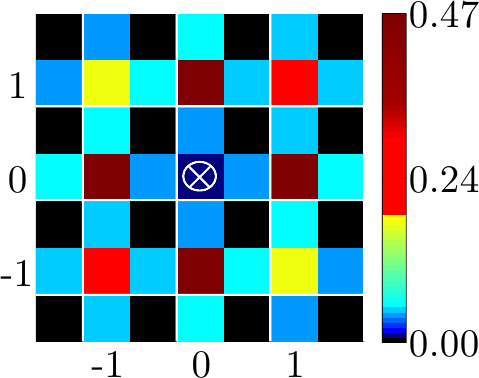}}%
\quad%
\subfloat[\textcolor{white}{h}]{\includegraphics[width=0.3\columnwidth]{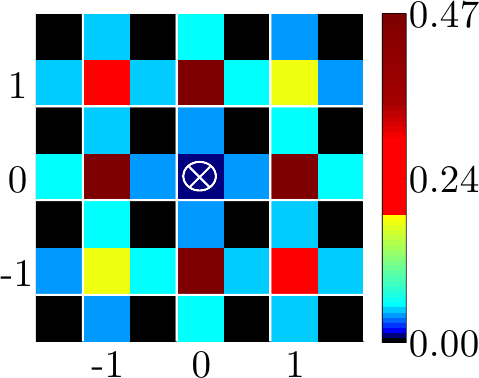}}%
\quad%
\subfloat[\textcolor{white}{l}]{\includegraphics[width=0.3\columnwidth]{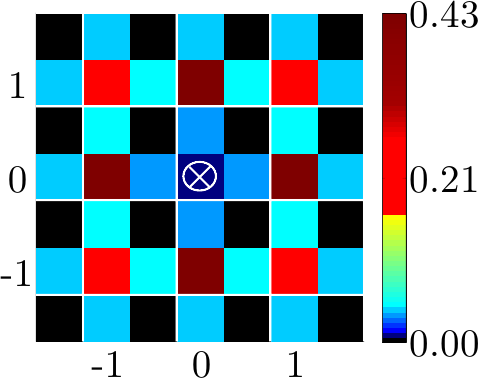}}%
\caption{$C_4$ symmetry breaking near an impurity positioned at the origin $(0,0)$ by loop currents. 
 (a)-(c) Possible loop current patterns involving a central Cu and four neighboring O orbitals.
(d)-(f) Corresponding LDOS at $\omega=0.2$ eV on the Cu and O orbitals near
a single nonmagnetic impurity at the central Cu site marked by a white cross. Black
  sites are vacant, numbers label Cu atoms.  Results are for (a),(d) ${\cal A}=1$,
(b),(e) ${\cal A}=2$, (c),(f) ${\cal A}=3$.}%
\label{fig:LDOSpatterns}
\end{figure} 

The LDOS patterns
in the vicinity of the impurity are shown in Fig.~\ref{fig:LDOSpatterns}, along with
the corresponding current patterns. The current pattern denoted by
$\mathcal{A}=1$ is generated by the Hamiltonian~\eqref{eq:3by3Hamiltonian}. The other two current patterns are obtained by changing
the off-diagonal entries of $H(\k)$:
\begin{align}
 \mathcal{A}=2:&\quad H_{12}(\k)=2it_{pd}s_x-Rc_x,\\
               &\quad H_{13}(\k)=-2it_{pd}s_y+Rc_y. \\
\mathcal{A}=3:&\quad H_{12}(\k)=2it_{pd}s_x+Rs_x,\\
               &\quad H_{13}(\k)=-2it_{pd}s_y+Rs_y. 
\end{align}

Schematically, the unit cell in real space can be represented as
\[
  \bmx p_y & 0 \\ d & p_x \emx,
\]
which also provides a legend for reading Fig.~\ref{fig:LDOSpatterns}. As can be seen from Fig.~\ref{fig:LDOSpatterns}, the symmetries of the LDOS patterns are inherited from the spatial symmetries of the loop order, with the obvious additional rule that the directions of the current arrows play no role in the LDOS.  
These symmetries are tabulated in Table~\ref{tab:numerical_symmetries}, and are easily obtained from the symmetries of Eq.~(\ref{Gfull}).
We stress that in the clean phase, the LDOS has the full symmetry of the lattice, and that it is only
near an impurity that the distinctive signatures of the loop currents
are revealed.

\begin{table}[t]
\centering
\begin{tabular}{|l|l|l|l|l|l|l|}
\hline
     Current pattern &  $\mathcal{A}=1$ & $\mathcal{A}=2$ & $\mathcal{A}=3$\\ \hline
Symmetry $\down$  & - & - & - \\ \hline
   $x$-axis refl. & No &   No &  Yes  \\ \hline
   $y$-axis refl. & No  &  No  & Yes \\ \hline
   inversion      & Yes  & Yes  & Yes \\ \hline
   $x=y$ refl.    & Yes  & Yes  & Yes  \\ \hline
   $x=-y$ refl.   & Yes  & Yes  & Yes \\ \hline
   C4             & No  & No  & Yes    \\ \hline
\end{tabular}
\caption{Symmetries of the LDOS near an impurity. 
}
\label{tab:numerical_symmetries}
\end{table}

\begin{figure}[b]
\centering
\subfloat[]{\includegraphics[width=0.23\textwidth]{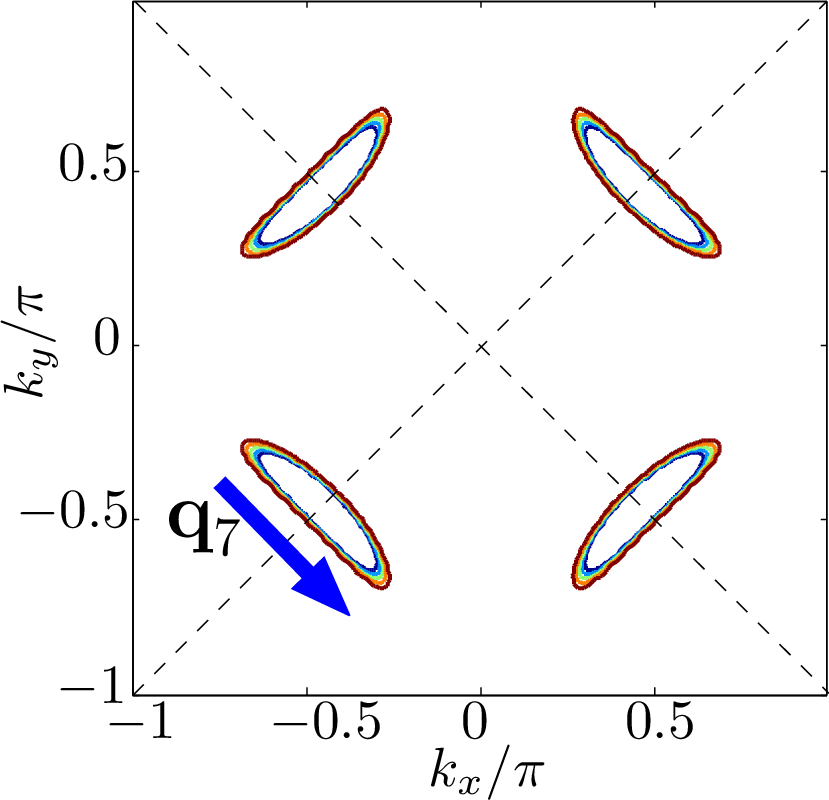}}
\quad
\subfloat[]{\includegraphics[width=0.23\textwidth]{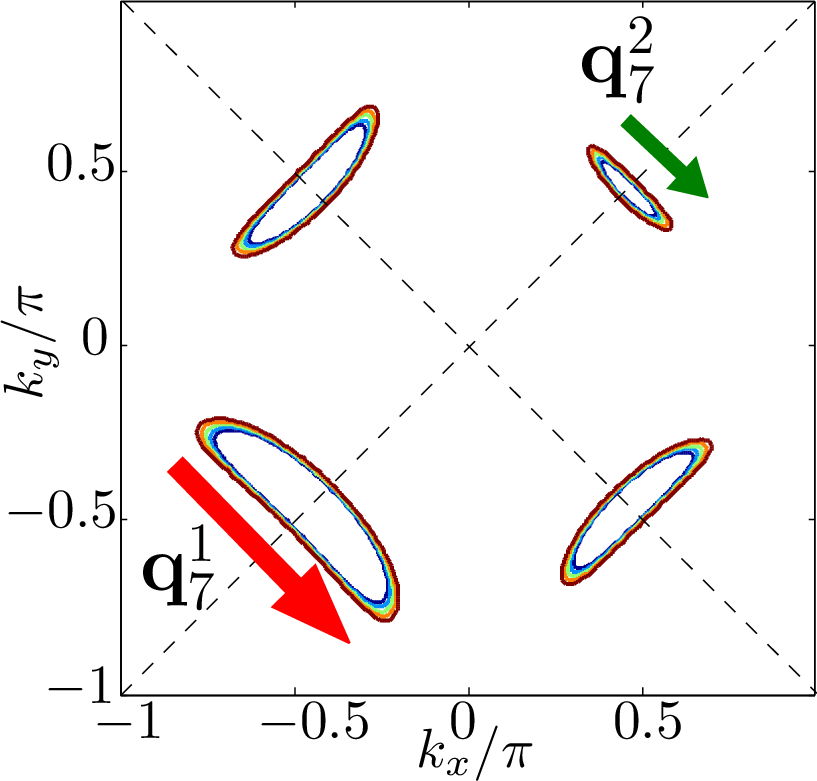}}
\caption{Contours of constant energy for energies near
$\omega=0.2$ eV (a) without and (b) with  $\mathcal{A}=1$ orbital loop order.}
\label{fig:bananas}
\end{figure}

We can understand the broken spatial symmetries presented in Fig.~\ref{fig:LDOSpatterns} by consulting the contours of constant energy (CCE) of the bare band structure. As in the case of a pure $d$-wave superconductor, the band structure inside the gap yields closed banana-shaped CCE which increase in size with the energy $\omega$ as shown in Fig.~\ref{fig:bananas}(a). In Fig.~\ref{fig:bananas}(b) we see the effect of orbital currents with $\mathcal{A}=1$ on the CCE; while two of the Dirac cones are left invariant, the other two are shifted upwards/downwards in energy.\cite{berg} 
These distorted bananas respect the same symmetries as the current patterns. Changing the current direction corresponds to interchanging the large and small bananas whereas changing the current pattern $\mathcal{A}=1 \rightarrow \mathcal{A}=2$ causes a $C_4$ rotation of the LDOS as expected. 

\begin{figure}[t]
\centering
\subfloat[]{\includegraphics[width=0.45\columnwidth]{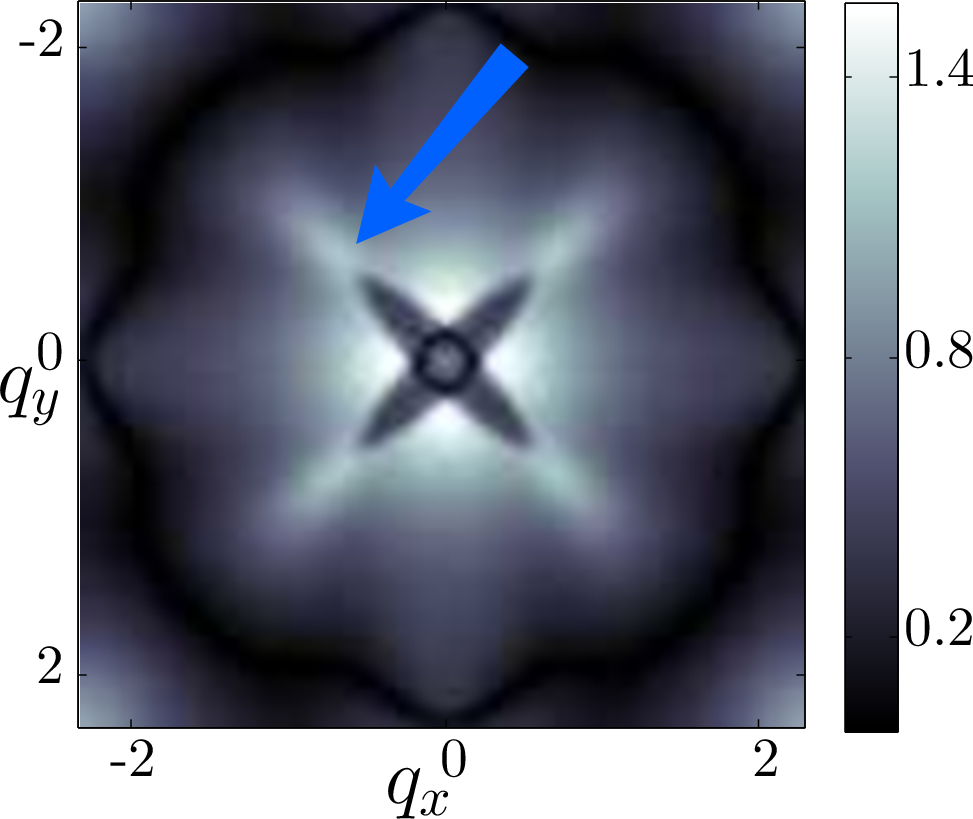}}
\quad
\subfloat[]{\includegraphics[width=0.45\columnwidth]{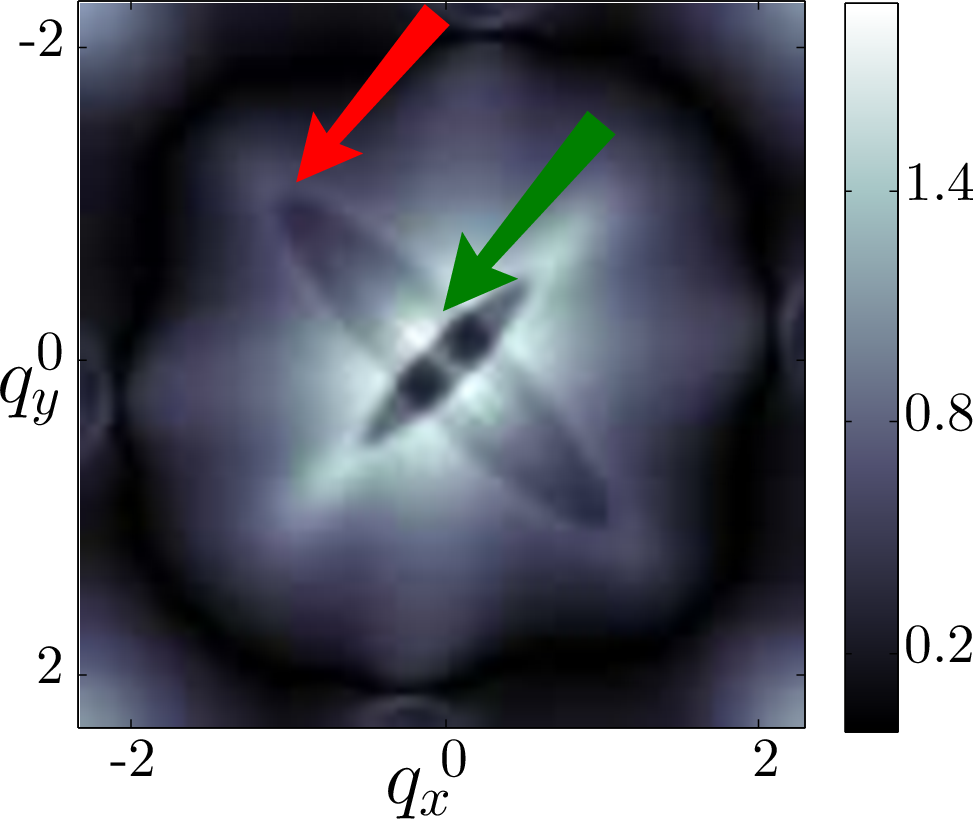}}

\subfloat[]{\includegraphics[width=0.45\columnwidth]{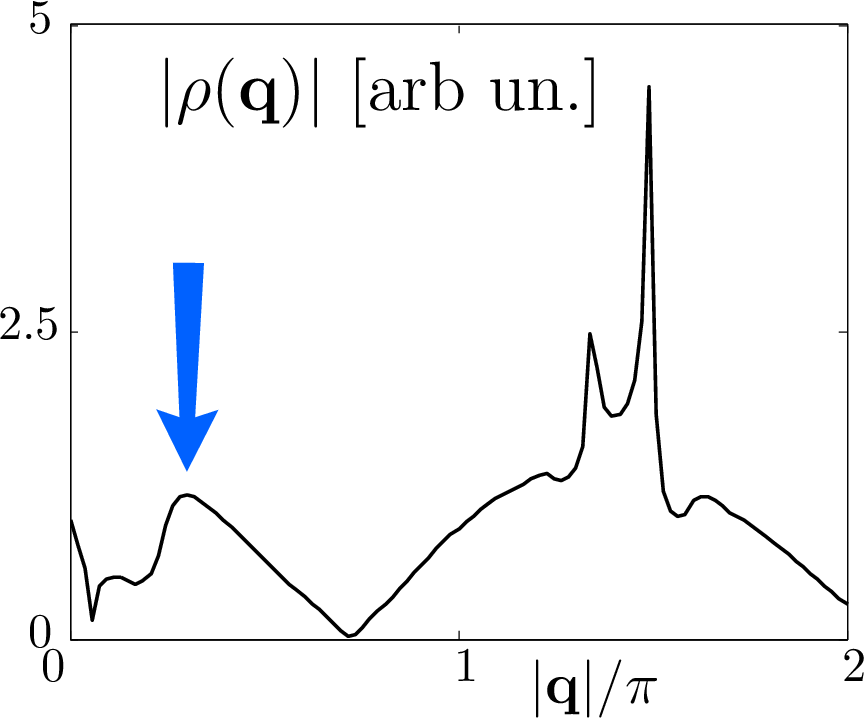}}
\quad
\subfloat[]{\includegraphics[width=0.45\columnwidth]{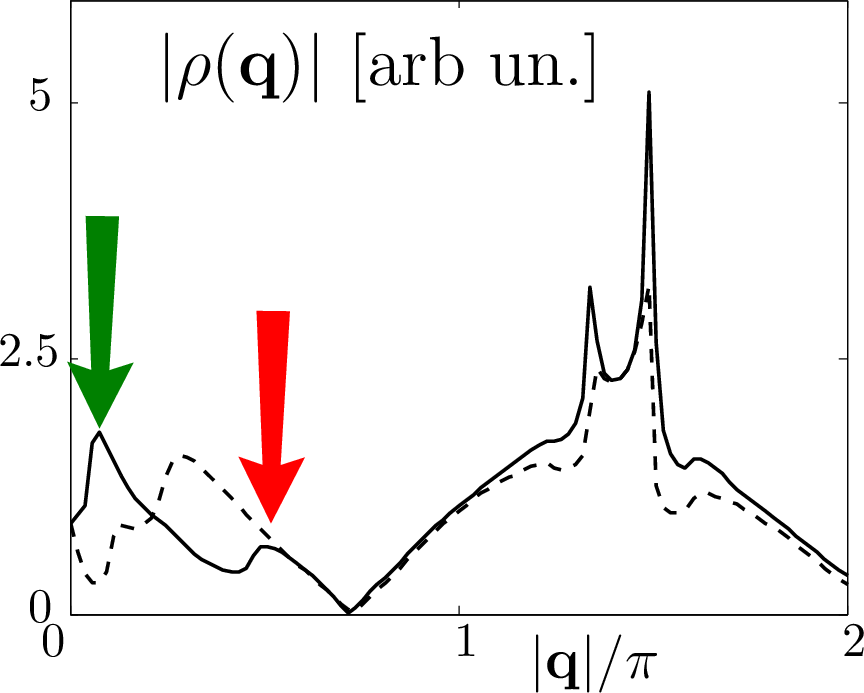}}

\subfloat[]{\includegraphics[width=0.65\columnwidth]{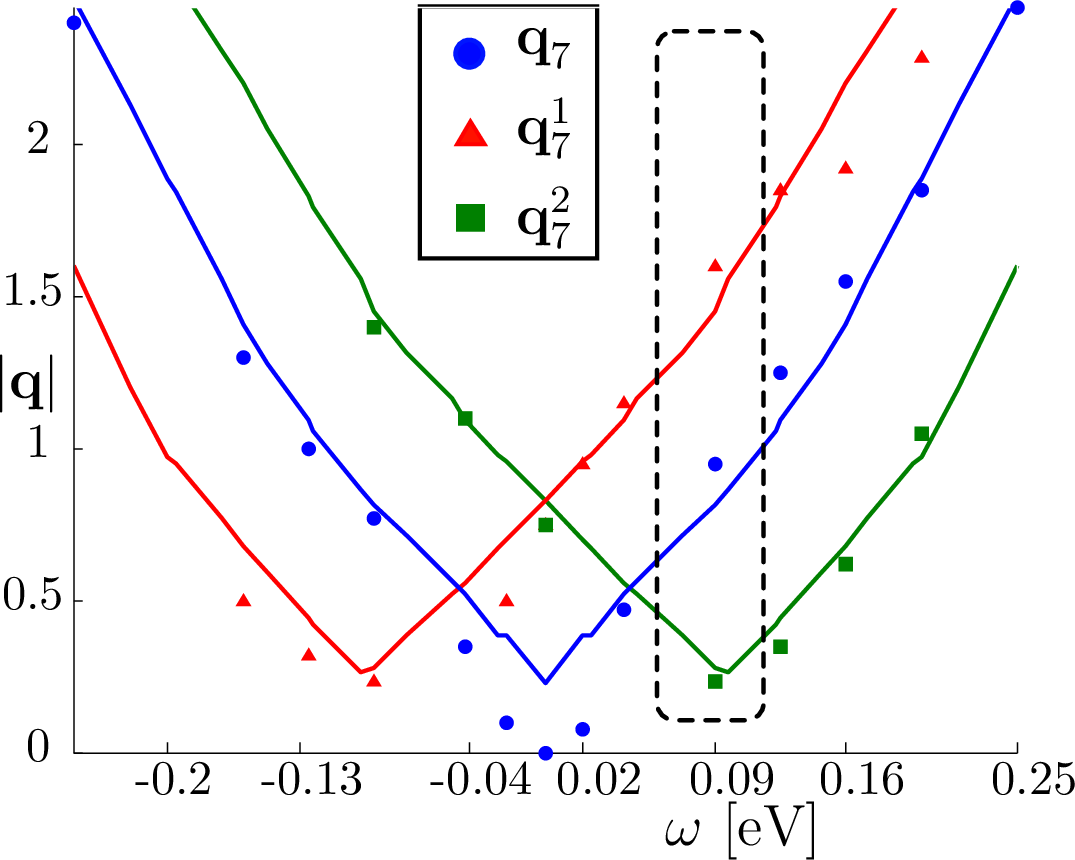}\label{fig:PaperFigure3e}}
 \caption{Plot of  $|\rho(\q,\omega)|$ at $\omega=0.09$ eV (a) without and (b) with ${\cal A}=1$ loop currents.   Arrows indicate (a) $\q_7$ and (b) $\q^1_7$ and $\q^2_7$ peaks. Cuts of $|\rho(\q,\omega)|$ are shown along $q_y=-q_x$ (solid) and $q_y=q_x$ (dashed) (c) without and (d) with loop currents.  Arrows indicate the same peaks as in (a) and (b).   (e) Dispersion of the $\q_7$ peaks as a function of $\omega$.   There is good agreement between the peak positions in $\rho(\q,\omega)$ (points) and the octet model (solid lines).}
\label{fig:PaperFigure3}
\end{figure}

Experimentally, the CCE may be inferred from QPI patterns measured by tunneling experiments. For 
pure $d$-wave superconductors, these QPI patterns have been explained using a simple  ``octet'' model,\cite{hoffman,wanglee,capriotti,nunner,andersen} in which peaks in the QPI pattern are attributed to scattering between
portions of the CCE for which the joint density of states is high. Figure~\ref{fig:bananas}(a) shows the octet vector $\q_7$, which labels the most prominent ``intra-banana'' quasiparticle scattering process.

The effect of ${\cal A}=1$ loop currents on $\q_7$ is illustrated in Fig.~\ref{fig:bananas}(b): the $\q_7$ peaks along the left diagonal are split into $\q^1_7$
and $\q^2_7$, while those along the right diagonal remain unchanged. This uniaxial distortion is certainly seen in the calculated QPI spectrum shown in Fig.~\ref{fig:PaperFigure3}(a-d) which display $|\rho(\q,\omega)|$ at $\omega=0.09$ eV as a function of $\q$. In addition,
Fig.~\ref{fig:PaperFigure3}(e) shows how the identification of the $\q_7$ peaks in the QPI spectrum is in very good agreement with the octet model prediction obtained directly from the CCEs. As is evident from e.g. Fig.~\ref{fig:bananas}(b), the other (than $\q_7$) characteristic $\q$-vectors of the octet model connecting the banana tips will be slightly rotated off the center axis in the presence of circulating current loop order. 

\section{Discussion and Conclusion}

There are a number of possible caveats to the present proposal that should be discussed. First of all, the presence of dynamical loop order fluctuations will clearly be detrimental to the observation of $C_4$ spatial symmetry breaking. Second, the existence of various domains with respectively $\mathcal{A}=1$ and $\mathcal{A}=2$ loop order will "symmetrize" the QPI images and roughly result in a smeared version of the QPI from a pure $d$-wave superconductor without current loops. The local $C_4$ symmetry breaking near a single impurity should still be detectable in this case. For BSCCO, however, it is expected that the LDOS near single-site disorder is further complicated by intrinsic disorder and the structural supermodulation.\cite{hamidian,andersenZn} Finally, we should also mention that the current calculation is not self-consistent and the feedback effect of the impurity on the local current loops is not included.  The effect of a charged impurity on the loops has been addressed previously in the discussion of the potential disturbance of the $\mu^+$  on the current order.\cite{shekter,millis} In the present case it is unlikely that the local suppression of loop currents would restore the symmetry. Thus, we expect the overall conclusion of $C_4$ symmetry breaking near disorder sites to remain valid.  

In summary, we have shown how spatial $C_4$ symmetry may become broken in the LDOS around nonmagnetic impurities in the coexistence phase of loop order and $d$-wave superconductivity. In addition there exists an associated splitting and/or rotation of the so-called octet peaks evident in the quasiparticle interference patterns. These predictions should be testable by future tunneling measurements. 

\section{Acknowledgements}

We acknowledge useful discussions with J.~C.~Davis. B.~M.~A.\ acknowledges support from The Danish Council for Independent Research $|$ Natural Sciences. W.~A.~A.\ acknowledges support from NSERC of Canada.


\begin{thebibliography} {00}

\bibitem{fauque} B. Fauqu\'{e}, Y. Sidis, V. Hinkov, S. Pailh\'{e}s, C. T. Lin, X. Chaud, and P. Bourges, Phys. Rev. Lett. {\bf 96}, 197001 (2006). 
%
\bibitem{mook} H. A. Mook, Y. Sidis, B. Fauqu\'{e}, V. Bal\'{e}dent, and P. Bourges, Phys. Rev. B  {\bf 78}, 020506(R) (2008).
%
\bibitem{li} Y. Li, V. Bal\'{e}dent, N. Barisi\'{c}, Y. C. Cho, B. Fauqu\'{e}, Y. Sidis, G. Yu, X. Zhao, P. Bourges, and M. Greven, Nature (London) {\bf 455}, 372 (2008). 
%
\bibitem{li2} Y. Li, V. Bal\'{e}dent, N. Barisi\'{c}, Y. C. Cho, Y. Sidis, G. Yu, X. Zhao, P. Bourges, and M. Greven, Phys. Rev. B {\bf 84}, 224508 (2011).
%
\bibitem{baledent} V. Bal\'{e}dent, B. Fauqu\'{e}, Y. Sidis, N. B. Christensen, S. Pailh\'{e}s, K. Conder, E. Pomjakushina, J. Mesot, and P. Bourges, Phys. Rev. Lett. {\bf 105}, 027004 (2010).
%
\bibitem{varma} C. M. Varma, Phys. Rev. Lett. {\bf 83}, 3538 (1999).
%
\bibitem{varma2} C. M. Varma, Phys. Rev. B {\bf 73}, 155113 (2006).
%
\bibitem{weber} C. Weber, A. L\"{a}uchli, F. Mila, and T. Giamarchi, Phys. Rev. Lett. {\bf 102}, 017005 (2009).
%
\bibitem{straassle_old} S. Str\"{a}ssle, J. Roos, M. Mali, H. Keller, and T. Ohno, Phys. Rev. Lett.  {\bf 101}, 237001 (2008).
%
\bibitem{macdougall} G.J. MacDougall, A. A. Aczel, J. P. Carlo, T. Ito, J. Rodriguez, P. L. Russo, Y. J. Uemura, S. Wakimoto, and G. M. Luke, Phys. Rev. Lett.  {\bf 101}, 017001 (2008).
%
\bibitem{sonier} J. E. Sonier, V. Pacradouni, S. A. Sabok-Sayr, W. N. Hardy, D. A. Bonn, R. Liang, and H. A. Mook, Phys. Rev. Lett. {\bf 103}, 167002 (2009).
%
\bibitem{straassle} S. Str\"{a}ssle, B. Graneli, M. Mali, J. Roos, and H. Keller, Phys. Rev. Lett. {\bf 106}, 097003 (2011).
%
\bibitem{huang} W. Huang, V. Pacradouni, M. P. Kennett, S. Komiya, and J. E. Sonier, ArXiv:1201.5406v1.
%
\bibitem{shekter} A. Shekhter, L. Shu, V. Aji, D. E. MacLaughlin, and C. M. Varma, Phys. Rev. Lett. {\bf 101}, 227004 (2008).
%
\bibitem{millis} H. T. Dang, E. Gull, and A. J. Millis, Phys. Rev. B {\bf 81}, 235124 (2010).
%
\bibitem{lawler} M. J. Lawler, K. Fujita, J. Lee, A. R. Schmidt, Y. Kohsaka,  C. K. Kim, H. Eisaki, S. Uchida, J. C.  Davis, J. P.  Sethna, and E.-A. Kim, Nature {\bf 466}, 346 (2010).
%
\bibitem{berg} E. Berg, C-C. Chen, and S. A. Kivelson, Phys. Rev. Lett. {\bf 100}, 027003 (2008).
%
\bibitem{chakravarty} S. Chakravarty, R.B. Laughlin, D.K. Morr, and C. Nayak, Phys. Rev. B {\bf 63}, 094503 (2001).
%
\bibitem{wang} Q.-H. Wang, Phys. Rev. Lett. {\bf 88}, 057002 (2002). 
%
\bibitem{morr} D. K. Morr, Phys. Rev. Lett. {\bf 89}, 106401 (2002).
%
\bibitem{andersenddw} B. M. Andersen, Phys. Rev. B {\bf 68}, 094518 (2003).
%
\bibitem{fischer} M. H. Fischer and E.-A. Kim, Phys. Rev. B {\bf 84}, 144502 (2011).
%
\bibitem{hybertsen} M. S. Hybertsen, E. B. Stechel, M. Schluter, and D. R. Jennison, Phys. Rev. B {\bf 41}, 11068 (1990). 
%
\bibitem{footnote} Experimentally, the magnetic elastic neutron signal corresponds to a dipole moment of $\sim 0.05 \mu_\mathrm{B}$ in underdoped YBa$_2$Cu$_3$O$_{6+x}$.\cite{fauque} For the mean-field model presented here, the particular value $R=0.1i$ generates a current of $j\sim 0.01 eV/\hbar$ which gives rise to a magnetic moment per unit cell of approximately $\sim0.01 \mu_B$. The current $j$ obtained from $R$ is, however, band-structure dependent, and we find that larger values of $R$ strongly modifies the low-energy LDOS in ways that seem inconsistent with STM data on BSCCO. 
%
\bibitem{hoffman} J. E. Hoffman, K. McElroy, D.-H. Lee, K. M Lang, H. Eisaki, S. Uchida, and J. C. Davis, Science {\bf 297}, 1148 (2002).
%
\bibitem{wanglee} Q.-H. Wang and D.-H. Lee, Phys. Rev. B {\bf 67}, 020511 (2003). 
%
\bibitem{capriotti} L. Capriotti, D. J. Scalapino, and R. D. Sedgewick, Phys. Rev. B {\bf 68}, 014508 (2003).
%
\bibitem{nunner} T. S. Nunner, W. Chen, B. M. Andersen, A. Melikyan, and P. J. Hirschfeld, Phys. Rev. B {\bf 73}, 104511 (2006).
%
\bibitem{andersen} B. M. Andersen and P. J. Hirschfeld, Phys. Rev. B {\bf 79}, 144515 (2009).
%
\bibitem{hamidian} M. H. Hamidian, I. A. Firmo, K. Fujita, S. Mukhopadhyay, J. W. Orenstein, H. Eisaki, S.-I. Uchida, M. J. Lawler, E.-A. Kim, J. C. Davis, 	arXiv:1202.4320v1.
%
\bibitem{andersenZn} B. M. Andersen, S. Graser, and P. J. Hirschfeld, Phys. Rev. B {\bf 78}, 134502 (2008).
 
\end{thebibliography}
\end{document}